# MINING TARGET-ORIENTED SEQUENTIAL PATTERNS WITH TIME-INTERVALS

Hao-En Chueh

Department of Information Management, Yuanpei University, Hsinchu City, Taiwan
hechueh@mail.ypu.edu.tw

## ABSTRACT

*A target-oriented sequential pattern is a sequential pattern with a concerned itemset in the end of pattern. A time-interval sequential pattern is a sequential pattern with time-intervals between every pair of successive itemsets. In this paper we present an algorithm to discover target-oriented sequential pattern with time-intervals. To this end, the original sequences are reversed so that the last itemsets can be arranged in front of the sequences. The contrasts between reversed sequences and the concerned itemset are then used to exclude the irrelevant sequences. Clustering analysis is used with typical sequential pattern mining algorithm to extract the sequential patterns with time-intervals between successive itemsets. Finally, the discovered time-interval sequential patterns are reversed again to the original order for searching the target patterns.*

## KEYWORDS

*Data Mining, Target-Oriented Sequential Pattern, Time-Interval, Clustering Analysis*

## 1. INTRODUCTION

Data mining (sometimes called knowledge discovery in databases) is the process of finding correlations or patterns among dozens of fields in large relational databases [4, 9, 12]. The primary tasks of data mining include association analysis, clustering analysis, classification, pattern recognition, prediction, etc. It has been widely used in business and engineering fields to discover useful information but previously unknown [15, 16, 20, 23].

Discovering sequential patterns is one of the most important tasks in data mining. The procedure of data mining is focused on mining sequential patterns is the task of finding frequently occurring patterns related to time or other sequences from a sequence database [1]. An example of a sequential pattern is "A customer who bought a digital camera will buy an extra memory card later". This kind of data mining task is very useful in retail business to assist decision makers in making marketing strategies [4, 9, 12, 15, 23].

Up to now, many mining sequential patterns algorithms have been proposed [1, 3, 5, 10, 14, 15, 18, 21], and most algorithms only focus on the order of the itemsets, but ignore the time-intervals between itemsets. However, in retail business, a sequential pattern with time-intervals between itemsets is more valuable than a traditional sequential pattern without any time information. An example of a sequential pattern with time-interval between itemsets is "A customer who bought a digital camera will buy an extra memory card within one week". Clearly, the time-intervals between itemsets can offer useful information for retail business to provide the correct products to their customers at the right time. Therefore, some researches start to focus on discovering the sequential patterns with time-intervals between itemsets [2, 6, 8, 11, 13, 17, 18, 19, 23]. This kind of sequential pattern is called as time-interval sequential pattern.

To obtain time-intervals between every pair of successive itemsets, clustering analysis is used in this paper to automatically generate the suitable time partitions between frequent occurring pairs of successive itemsets, and then uses these obtained time-intervals to extend typical algorithms to discover the time-interval sequential patterns [8].

In addition, for most marketing decision makers, they usually need to know the happening order of some concerned itemsets. This kind of sequential pattern ca ne called as target-oriented sequential pattern [7]. To this end, the original sequences are reversed so that the last itemsets can be arranged in front of sequences. The contrasts between the reversed sequences and the concerned itemset are then used to exclude the irrelevant sequences. These reversed sequences are analyzed to extract sequential patterns with time-interval between successive itemsets. Finally, the discovered time-interval sequential patterns are reversed again to the original order for searching the target patterns.

An algorithm to discover target-oriented sequential pattern with time-intervals between successive itemsets is presented in this paper. The rest of this paper is organized as follows: Some researches related to target-oriented sequential patterns and time-interval sequential patterns are reviewed in section 2. The algorithm to mine target-oriented sequential patterns with time-intervals between successive itemsets is presented in section 3. A simple example is displayed in section 4. The conclusions are given in section 5.

## 2. TARGET-ORIENTED AND TIME-INTERVAL SEQUENTIAL PATTERNS

Sequential patterns mining was first introduced by Agrawal and Srikant in the mid 1990s [1]. The procedure of sequential patterns mining can be described as the task of discovering frequently occurring ordered patterns from a given sequence dataset.

A sequence is an ordered list of itemsets. Let $I = \{i_1, i_2, ......, i_m\}$ be a set of items, $S = < s_1, s_2, ......, s_k >$ is a sequence, where $s_i \subseteq I$ is called an itemset. The length of a sequence means the number of itemsets in this sequence, and a sequence contains $k$ itemsets is called a k-sequence. The support of a sequence $S$ is denoted by $supp(S)$ and means the percentage of total number of records containing sequence $S$ in the sequence dataset. If $supp(S)$ is greater than or equal to a predefined threshold, called minimal support, than sequence $S$ is considered as a frequent sequence and called a sequential pattern [1, 9, 12, 22].

Many sequential patterns mining algorithms have been proposed up to now [1, 3, 5, 10, 14, 15, 18, 21], and most algorithms focus on finding the frequently occurring order of the itemsets, but ignore the time-intervals between successive itemsets. The time-intervals between successive itemsets of the sequential pattern can offer useful information for retail businesses to know sell the customers' need at the right time. Therefore, some researchers turn to focus on discovering sequential patterns with various time information recently [2, 6, 8, 11, 13, 17, 18, 19, 23].

Srikant et al. [22] use three predefined restrictions, the maximum interval ($max-interval$), the minimum interval ($min-interval$), and the time window size ($window-size$) to discover sequential patterns with time-intervals, and the obtained sequential pattern is like $((A,B),(C,D))$. The $max-interval$ and the $min-interval$ are used to indicate the maximal and the minimal interval within subsequence respectively. In addition, the $window-size$ is used to indicate the interval among every pairs of successive subsequences. For example, assume that the $min-interval$ is set to 3 days, the $max-interval$ is set to 8 days, and $window-size$ is set to 20 days. That is, the time-interval between $A$ and $B$ lies in $[3, 8]$, the time-interval between $C$ and $D$ also lies in $[3, 8]$, and the time-interval between $(A,B)$ and $(C,D)$ lies in $[0, 20]$.

Mannila et al. [19] only utilize a predefined restriction, window width ($win$), to find frequent episodes in sequences of events, and the discovered episode is like $(A, B, C)$. The episode $(A, B, C)$ means that, in ($win$) days, $A$ occurs first, $B$ follows, and $C$ happens finally.

Wu et al. [23] also use a predefined restriction, window ($d$), to discover the sequential patterns with time-intervals. The discovered sequential pattern likes $(A, B, C)$. The sequential pattern $(A, B, C)$ means that $A$ occurs first, $B$ follows, and $C$ happens finally and the-intervals between $A$ and $B$, and between $B$ and $C$ are both within ($d$) days.

Chen et al. [2] use a predefined set of non-overlap time partitions to discover potential sequential patterns with time-intervals, and the discovered pattern is like $(A, I_1, B, I_2, C)$, where $I_1, I_2$ belong to the predefined non-overlap set of time partitions. Assume that $I_1$ denotes the time partition [1, 3], $I_2$ denotes the time partition [7, 10], then the sequential pattern $(A, I_1, B, I_2, C)$ means that $A, B$ and $C$ happen in this order, and the time-interval between $A$ and $B$ lies between 1 day and 3 days., and the time-interval between $B$ and $C$ lies between 7 days and 10 days.

Chiang et al. [6] propose a new method called ISP to find the periodic interval of consumer items of each customer. Compare with other previous researches, the difference is that the period the algorithm provides is not the repeated purchases in a regular time, but the possible repurchases within a certain time frame (shortest probable time, average probable time and longest probable time). The probability based algorithm utilizes the transaction time interval of individual customers and that of all the customers to find out when and who will buy goods, and what items of goods they will buy.

Although these previous researches can discover the sequential patterns with the time-intervals between itemsets, they may encounter the sharp boundary problem. Therefore, this work propose to use the clustering analysis to automatically generate the time-intervals for frequent occurring pairs of successive itemsets, and then use these generated time-intervals to extend typical algorithms to mine the sequential patterns with time-intervals between successive itemsets without defining any time partitions in advance [8].

Sequential pattern mining is widely used in retail business to assist decision makers in making marketing strategies. However, for most decision makers, when they want to make efficient marketing strategies, they usually concern the happening order of a concerned itemsets only, and thus, most sequential patterns discovered by using traditional algorithms are irrelevant and useless. To solve this problem, Chiang et al. [7] use the concept of "the reversed sequence" to implement the quick searches for particular itemset. First, the original sequences are reversed so that the last itemsets can be arranged in front of the sequences. Next, the reversed sequences are compared with the concerned itemsets to sift the irrelevant sequences. These retained sequences are analyzed to extract sequential patterns. Finally, the discovered sequential patterns are reversed again to the original order for searching the target patterns.

The main objective of this paper is to present an algorithm for discovering target-oriented sequential pattern with time-intervals between successive itemsets. Details of the proposed algorithm are presented in the next section.

## 3. TARGET-ORIENTED SEQUENTIAL PATTERNS WITH TIME-INTERVALS

The proposed algorithm for mining target-oriented time-interval sequential patterns is presented in this section. The main concept of this algorithm is to reverse the original sequences so that the last itemsets can be arranged in front of the sequences. Next, the clustering analysis is used with the traditional sequential patterns mining algorithms to discover the sequential patterns with time-intervals. Finally, each discovered time-interval sequential pattern is re-reversed again to return the original order. First, some notations are explained as follows:

$S_i = <s_1, s_2, ......, s_n>$ : A sequence, where $s_i$ is an itemset.
$D = \{S_1, S_2, ......, S_k\}$ : The sequence dataset.
$supp(S_i)$ : The support of $S_i$.
$min-supp$ : The minimal support threshold.
$CS_k$ : The set of candidate frequent k-sequences.
$FS_k$ : The set of frequent k-sequences.
$CTIS_k$ : The set of candidate frequent time-interval k-sequences.
$FTIS_k$ : The set of frequent time-interval k-sequences.

Next, the detailed procedure of the proposed algorithm is described in the next subsection.

### 3.1. The Proposed Algorithm:

Step 1: Reverse each original sequence so that the last itemset of the sequence can be arranged in front of the sequence.

Step 2: Remove the reversed sequences without the concerned itemsets.

Step 3: Delete itemsets which are arranged in front of the concerned itemsets from each retained sequences.

Step 4: Produce $CS_1$, the set of candidate frequent 1-sequences. Each itemset in the changed sequence dataset can be regarded as a candidate frequent 1-sequence.

Step 5: Produce $FS_1$, the set of frequent 1-sequences. A member of $CS_1$ whose support is greater than or equal to $min-supp$ is a frequent 1-sequence, and the set of all frequent 1-sequences is $FS_1$.

Step 6: Produce $CS_2$, the set of candidate frequent 2-sequences. For any two frequent 1-sequences $<s_1>$ and $<s_2>$, where $<s_1>, <s_2> \in FS_1$ and $s_1 \neq s_2$, then we can generate 2 sequences $<s_1, s_2>$ and $<s_2, s_1>$ that belong to $CS_2$.

Step 7: Produce $FS_2$, the set of frequent 2-sequences. A member of $CS_2$ whose support is greater than or equal to $min-supp$ is a frequent 2-sequence, and the set of all frequent 2-sequences is $FS_2$.

Step 8: Find the frequent time-intervals for each frequent 2-sequence. For any member of $FS_2$, say $<s_p, s_q>$, list all the time-intervals between $s_p$ and $s_q$ appear in $D$ in increasing order, then use the following clustering analysis steps, step 8(a), step 8(b) and step 8(c) to find all frequent time-intervals of $<s_p, s_q>$.

Step 8(a): Assume that $[t_1, t_2, \cdots, t_z]$ is the increasingly ordered list of the time-intervals of $<s_p, s_q>$. Let $TI<s_p, s_q> = \{[t_1, t_2, \cdots, t_z]\}$ be the set of time-intervals of $<s_p, s_q>$. The first step of a clustering analysis is to find the maximal difference between two adjacent time-intervals of $TI < s_p, s_q >$, and then divide $TI < s_p, s_q >$ into 2 subsets according to the maximal difference. Assume that the difference between $t_i$ and $t_{i+1}$ is maximal, then $TI < s_p, s_q >$ is divided into $[t_1, \cdots, t_i]$ and $[t_{i+1}, \cdots, t_z]$.

Step 8(b): Calculate the supports of $< s_p, s_q >$ with $[t_1, \cdots, t_i]$ and with $[t_{i+1}, \cdots, t_z]$ respectively If the support of $< s_p, s_q >$ with $[t_1, \cdots, t_i]$ is greater than or equal to $min-supp$, then time-intervals set $[t_1, \cdots, t_i]$ is a frequent time-interval of $< s_p, s_q >$, and then reserve $[t_1, \cdots, t_i]$, otherwise delete $[t_1, \cdots, t_i]$. Similarly, if the support of $< s_p, s_q >$ with $[t_{i+1}, \cdots, t_z]$ is greater than or equal to $min-supp$, then time-intervals set $[t_{i+1}, \cdots, t_z]$ is also a frequent time-interval of $< s_p, s_q >$, and reserve $[t_{i+1}, \cdots, t_z]$, otherwise delete $[t_{i+1}, \cdots, t_z]$. Next, the reserved time-intervals subsets replace the original time-interval set $TI < s_p, s_q >$. If no subset is reserved, then the original time-interval set is a non-dividable set. If all differences between two adjacent time-intervals are equal, then the original time-interval set is a non-dividable set as well.

Step 8(c): Repeat step 8(a) and step 8(b), until all subsets of time-intervals in $TI < s_p, s_q >$ are non-dividable sets.

Step 9: Produce $FTIS_2$, the set of frequent time-interval 2-sequences. Extend each member of $FS_2$ by using all its frequent time-intervals to generate frequent time-interval 2-sequence. If $TI < s_p, s_q >= \{T_{pq}^1, T_{pq}^2, \cdots, T_{pq}^r\}$ is the set of frequent time-intervals of the sequence $< s_p, s_q >$, then all $< s_p, T_{pq}^i, s_q >, i=1 \cdots r$, are frequent time-interval 2-sequences. The set of frequent time-interval 2-sequences is $FTIS_2$.

Step 10: Produce $CTIS_k, k \geq 3$, the set of candidate frequent time-interval k-sequences. For any two frequent time-interval (k-1)-sequences $S_1$ and $S_2$, where $S_1 =< s_{1,1}, T_{1,1}, s_{1,2}, \cdots, s_{1,k-2}, T_{1,k-2}, s_{1,k-1} >$; $S_2 =< s_{2,1}, T_{2,1}, s_{2,2}, \cdots, s_{2,k-2}, T_{2,k-2}, s_{2,k-1} >\in FTIS_{k-1}$; $s_{1,2} = s_{2,1}, s_{1,3} = s_{2,2}, \cdots, s_{1,k-1} = s_{2,k-2}$; $T_{1,2} = T_{2,1}, T_{1,3} = T_{2,2}, \cdots, T_{1,k-2} = T_{2,k-3}$, then we can generate a candidate frequent time-interval k-sequences $S_1 =< s_{1,1}, T_{1,1}, s_{1,2}, \cdots\cdots, s_{1,k-2}, T_{1,k-2}, s_{1,k-1}, T_{2,k-2}, s_{2,k-1} >$. The set of candidate frequent time-interval k-sequences is $CTIS_k$.

Step 11: Produce $FTIS_k, k \geq 3$, the set of frequent time-interval k-sequences. A member of $CTIS_k$ whose support is greater than or equal to $min-supp$ is a frequent time-interval k-sequence, and the set of frequent time-interval k- sequences is $FTIS_k$.

Step 12: Repeat step 10 and step 11, until no next $CTIS_{k+1}$ can be generated.

Step 13: Remove the frequent time-interval sequences without the concerned itemset.

Step 14: Reverse the retained sequences again to return the original order.

In the next section, a simple example is used to illustrate the procedure of mining target-oriented time-interval sequential patterns.

## 4. EXAMPLE

In this section, we use the sample sequence databset shown as in Table 1 to discover the target-oriented sequential patterns with time-intervals. In Table 1, CNo denotes the customer number of a sequence, and each sequence is represented as $<(s1,t_1),(s2,t_2),\cdots,(sn,t_n)>$, where $s_i$ denotes an itemset, and $t_i$ denotes the time stamp that $si$ occurs.

Table 1. Sample sequence dataset.

| CNo | Sequence |
|---|---|
| C001 | ($s5$,8), ($s4$,15), ($s6$,20) |
| C002 | ($s1$,2), ($s3$,7), ($s2$,11), ($s6$,18) |
| C003 | ($s2$,3), ($s1$,4), ($s3$,7), ($s6$,17), ($s7$,19) |
| C004 | ($s1$,2), ($s2$,8), ($s6$,10), ($s7$,15) |
| C005 | ($s5$,4), ($s6$,16), ($s1$,20), ($s3$,24) |
| C006 | ($s7$,7), ($s1$,13), ($s5$,18), ($s2$,25), ($s6$,28) |
| C007 | ($s5$,4), ($s1$,8), ($s3$,12), ($s6$,16), ($s7$,20) |
| C008 | ($s1$,3), ($s5$,6), ($s2$,9), ($s4$,18), ($s6$,21) |
| C009 | ($s2$,5), ($s1$,10), ($s3$,15), ($s6$,20), ($s7$,25) |
| C010 | ($s6$,3), ($s7$,8), ($s5$,12), ($s2$,17) |

According to section 3, the first step is to reverse each original sequence so that the last itemset of the sequence can be arranged in front of the sequence.

The next step is to remove the reversed sequences without the concerned itemsets. Here, assume that the concerned itemset is $s7$, and thus, the sequences of C0001, C0002, C0005 and C0008 are removed.

Table 2. The reversed sample sequence dataset.

| CNo | Sequence |
|---|---|
| C001 | ($s6$,20), ($s4$,15), ($s5$,8) |
| C002 | ($s6$,18), ($s2$,11) , ($s3$,7), ($s1$,2) |
| C003 | ($s7$,19), ($s6$,17), ($s3$,7), ($s1$,4), ($s2$,3) |
| C004 | ($s7$,15), ($s6$,10), ($s2$,8), ($s1$,2) |
| C005 | ($s3$,24), ($s1$,20), ($s6$,16), ($s5$,4) |
| C006 | ($s6$,28), ($s2$,25), ($s5$,18), ($s1$,13), ($s7$,7) |
| C007 | ($s7$,20), ($s6$,16), ($s3$,12), ($s1$,8), ($s5$,4) |
| C008 | ($s6$,21), ($s4$,18), ($s2$,9), ($s5$,6), ($s1$,3) |
| C009 | ($s7$,25), ($s6$,20), ($s3$,15), ($s1$,10), ($s2$,5) |
| C010 | ($s2$,17), ($s5$,12), ($s7$,8), ($s6$,3) |

Table 3. The sequence dataset with the concerned itemset $s7$.

| CNo | Sequence |
|---|---|
| C003 | ($s7$,19), ($s6$,17), ($s3$,7), ($s1$,4), ($s2$,3) |
| C004 | ($s7$,15), ($s6$,10), ($s2$,8), ($s1$,2) |
| C006 | ($s6$,28), ($s2$,25), ($s5$,18), ($s1$,13), ($s7$,7) |
| C007 | ($s7$,20), ($s6$,16), ($s3$,12), ($s1$,8), ($s5$,4) |
| C009 | ($s7$,25), ($s6$,20), ($s3$,15), ($s1$,10), ($s2$,5) |
| C010 | ($s2$,17), ($s5$,12), ($s7$,8), ($s6$,3) |

For each retained sequences, the itemsets which are arranged in front of the concerned itemset, $s7$, are deleted. Thus, sequences of C0006 and C0010 are changed.

Table 4. The sequence dataset with concerned itemset $s7$ in the first order.

| CNo | Sequence |
|---|---|
| C003 | ($s7$,19), ($s6$,17), ($s3$,7), ($s1$,4), ($s2$,3) |
| C004 | ($s7$,15), ($s6$,10), ($s2$,8), ($s1$,2) |
| C006 | ($s7$,7) |
| C007 | ($s7$,20), ($s6$,16), ($s3$,12), ($s1$,8), ($s5$,4) |
| C009 | ($s7$,25), ($s6$,20), ($s3$,15), ($s1$,10), ($s2$,5) |
| C010 | ($s7$,8), ($s6$,3) |

Next, the set of candidate frequent 1-sequences, $CS_1$, is recorded and listed. Each itemset in the changed sequence dataset is regarded as a candidate frequent 1-sequence, and thus, $CS_1 = \{<s1>, <s2>, <s3>, <s5>, <s6>, <s7>\}$. A member of $CS_1$ whose support is greater than or equal to $min-supp$ is a frequent 1-sequence. Here, $min-supp$ is set as 0.3. The set of all frequent 1-sequences is $FS_1$, $FS_1 = \{<s1>, <s2>, <s3>, <s6>, <s7>\}$.

Table 5. The supports of candidate frequent 1-sequences.

| $CS_1$ | Supp |
|---|---|
| $<s1>$ | 0.67 |
| $<s2>$ | 0.5 |
| $<s3>$ | 0.5 |
| $<s5>$ | 0.17 |
| $<s6>$ | 0.83 |
| $<s7>$ | 1 |

The set of candidate frequent 2-sequences, $CS_2$, is generated by joining $FS_1$ and itself. For any two frequent 1-sequences of $FS_1$, say $<s_1>$ and $<s_2>$, 2 candidate frequent 2-sequences, $<s_1, s_2>$ and $<s_2, s_1>$, can be generated. Thus, $CS_2 = \{<s1,s2>, <s1,s3>, <s1,s6>, <s1,s7>, <s2,s1>, <s2,s3>, <s2,s6>, <s2,s7>, <s3,s1>, <s3,s2>, <s3,s6>, <s3,s7>, <s6,s1>, <s6,s2>, <s6,s2>, <s6,s3>, <s6,s7>, <s7,s1>, <s7,s2>, <s7,s3>, <s7,s6>\}$. A member of $CS_2$ whose support is greater than or equal to $min-supp$ is a frequent 2-sequence, and the set of all frequent 2-sequences is $FS_2$. $FS_2 = \{<s1,s2>, <s3,s1>, <s3,s2>, <s6,s1>, <s6,s2>, <s6,s3>, <s7,s1>, <s7,s2>, <s7,s3>, <s7,s6>\}$.

Table 6. The supports of candidate frequent 2-sequences.

| $CS_2$ | Supp | $CS_2$ | Supp |
|---|---|---|---|
| $<s1,s2>$ | 0.33 | $<s3,s6>$ | 0 |
| $<s1,s3>$ | 0 | $<s3,s7>$ | 0 |
| $<s1,s6>$ | 0 | $<s6,s1>$ | 0.67 |
| $<s1,s7>$ | 0 | $<s6,s2>$ | 0.5 |
| $<s2,s1>$ | 0.17 | $<s6,s3>$ | 0.5 |
| $<s2,s3>$ | 0 | $<s6,s7>$ | 0 |
| $<s2,s6>$ | 0 | $<s7,s1>$ | 0.67 |
| $<s3,s7>$ | 0 | $<s7,s2>$ | 0.5 |
| $<s3,s1>$ | 0.5 | $<s7,s3>$ | 0.5 |
| $<s3,s2>$ | 0.33 | $<s7,s6>$ | 0.83 |

The following step is to find the frequent time-intervals for each frequent 2-sequence. All time-intervals of each member of $FS_2$ are listed in Table 7. According to the section 3, all frequent time-intervals for each frequent 2-sequence are obtained and listed in Table 8.

Table 7. The time-intervals of each member of $FS_2$.

| $FS_2$ | Time-intervals |
|---|---|
| $<s1, s2>$ | [1, 5] |
| $<s3, s1>$ | [3, 4, 5] |
| $<s3, s2>$ | [4, 10] |
| $<s6, s1>$ | [8, 8, 10, 13] |
| $<s6, s2>$ | [2, 14, 15] |
| $<s6, s3>$ | [4, 5, 10] |
| $<s7, s1>$ | [12, 13, 15, 15] |
| $<s7, s2>$ | [7, 16, 20] |
| $<s7, s3>$ | [8, 10, 12] |
| $<s7, s6>$ | [2, 4, 5, 5, 5] |

The next step is to extend each member of $FS_2$ by using all its frequent time-intervals to generate the set of frequent time-interval 2-sequences, $FTIS_2$. Thus, $FTIS_2 = \{<s1, T_{1,2}, s2>,$ $<s3, T_{3,1}, s1>, <s3, T_{3,2}, s2>, <s6, T_{6,1}, s1>, <s6, T_{6,2}, s2>, <s6, T_{6,3}, s3>, <s7, T_{7,1}^1, s1>, <s7, T_{7,1}^2, s1>, <s7, T_{7,2}, s2>, <s7, T_{7,3}, s3>, <s7, T_{7,6}, s6>\}$.

Table 8. The time-intervals of each member of $FS_2$.

| $FS_2$ | Frequent time-intervals |
|---|---|
| $<s1, s2>$ | $TI<s1, s2> = \{T_{1,2} = [1, 5]\}$ |
| $<s3, s1>$ | $TI<s3, s1> = \{T_{3,1} = [3, 4, 5]\}$ |
| $<s3, s2>$ | $TI<s3, s2> = \{T_{3,2} = [4, 10]\}$ |
| $<s6, s1>$ | $TI<s6, s1> = \{T_{6,1} = [8, 8]\}$ |
| $<s6, s2>$ | $TI<s6, s2> = \{T_{6,2} = [14, 15]\}$ |
| $<s6, s3>$ | $TI<s6, s3> = \{T_{6,3} = [4, 5]\}$ |
| $<s7, s1>$ | $TI<s7, s1> = \{T_{7,1}^1 = [12, 13], T_{7,1}^2 = [15, 15]\}$ |
| $<s7, s2>$ | $TI<s7, s2> = \{T_{7,2} = [16, 20]\}$ |
| $<s7, s3>$ | $TI<s7, s3> = \{T_{7,3} = [8, 10, 12]\}$ |
| $<s7, s6>$ | $TI<s7, s6> = \{T_{7,6} = [5, 5, 5]\}$ |

$CTIS_3$, the set of candidate frequent time-interval 3-sequences is generated by jointing $FTIS_2 \times FTIS_2$ Supports of the members of $CTIS_3$ are calculated and shown in Table 9. A member of $CTIS_3$ whose support is greater than or equal to $min-supp$ is called as a frequent time-interval 3-sequence. Therefore, we can obtain the set of the frequent time-interval 3-sequences, $FTIS_3 = \{<s3, T_{3,1}, s1, T_{1,2}, s2>, <s6, T_{6,3}, s3, T_{3,1}, s1>, <s7, T_{7,1}^2, s1, T_{1,2}, s2>, <s7, T_{7,3}, s3, T_{3,1}, s1>, <s7, T_{7,3}, s3, T_{3,2}, s2>\}$.

$CTIS_4$, the set of candidate frequent time-interval 4-sequences, is generated by jointing $FTIS_3 \times FTIS_3$. Here, only two candidate sequences, $<s6, T_{6,3}, s3, T_{3,1}, s1, T_{1,2}, s2>$ and $<s7, T_{7,3}, s3, T_{3,1},$ $s1, T_{1,2}, s2>$, are generated. The support of $<s6, T_{6,3}, s3, T_{3,1}, s1, T_{1,2}, s2>$ and the support of $<s7, T_{7,3}, s3, T_{3,1}, s1, T_{1,2}, s2>$ are both 0.33, thus, $<s6, T_{6,3}, s3, T_{3,1}, s1, T_{1,2}, s2>$ and $<s7, T_{7,3}, s3, T_{3,1}, s1, T_{1,2}, s2>$ are both frequent time-interval 4-sequences. Therefore, we can obtain the set of the frequent time-interval 4-sequences, $FTIS_4 = \{<s6, T_{6,3}, s3, T_{3,1}, s1, T_{1,2}, s2>, <s7, T_{7,3}, s3, T_{3,1}, s1, T_{1,2}, s2>\}$.

Table 9. The supports of candidate frequent 1-sequences.

| $CTIS_3$ | Supp |
|---|---|
| $<s3, T_{3,1}, s1, T_{1,2}, s2>$ | 0.33 |
| $<s6, T_{6,1}, s1, T_{1,2}, s2>$ | 0.17 |
| $<s6, T_{6,3}, s3, T_{3,1}, s1>$ | 0.33 |
| $<s6, T_{6,3}, s3, T_{3,2}, s2>$ | 0.17 |
| $<s7, T_{7,1}^1, s1, T_{1,2}, s2>$ | 0 |
| $<s7, T_{7,1}^2, s1, T_{1,2}, s2>$ | 0.33 |
| $<s7, T_{7,3}, s3, T_{3,1}, s1>$ | 0.5 |
| $<s7, T_{7,3}, s3, T_{3,2}, s2>$ | 0.33 |
| $<s7, T_{7,6}, s6, T_{6,1}, s1>$ | 0.17 |
| $<s7, T_{7,6}, s6, T_{6,2}, s2>$ | 0.17 |
| $<s7, T_{7,6}, s6, T_{6,3}, s3>$ | 0.17 |

No next $CTIS_5$ can be generated, thus the procedure stops here. The following step is to remove the frequent time-interval sequences without the concerned itemsets from $FTIS_i, i = 2 \cdots 4$. Finally, reverse the retained sequences again to return the original order. Thus, we can obtain the following frequent target-oriented time-interval sequences with the concerned itemset $s7$: $<s1, T_{7,1}^1, s7>, <s1, T_{7,1}^2, s7>, <s2, T_{7,2}, s7>, <s3, T_{7,3}, s7>, <s6, T_{7,6}, s7>, <s2, T_{1,2}, s1, T_{7,1}^2, s7>, <s1, T_{3,1}, s3, T_{7,3}, s7>, <s2, T_{3,2}, s3, T_{7,3}, s7>, <s2, T_{1,2}, s1, T_{3,1}, s3, T_{7,3}, s7>\}$. According to these discovered patterns, we can clearly see that the time-intervals between successive itemsets of every frequent target-oriented time-interval sequences are different and overlap; therefore, it is reasonable to generate the time-interval directly from the real sequence dataset when mining tasks. In addition, the experimental result shows that less irrelevant patterns are found in the mining procedure by using the proposed algorithm.

## 5. CONCLUSIONS

Traditional sequential pattern mining only focuses reveal the order of itemsets, but a sequential pattern with the time intervals between successive itemsets is more valuable than a traditional sequential pattern. In addition, when mining sequential patterns, most users usually concern about the happening order of some particular itemsets only, and thus, most sequential patterns discovered by using traditional algorithms are irrelevant and useless. To solving these two problems, a new algorithm for discovering target-oriented time-interval sequential patterns is presented in this paper. The main concept of this algorithm is to reverse the original sequence dataset to enhance the efficiency for searching the target patterns. In addition, the clustering analysis is used to automatically generate the suitable time partitions between successive itemsets so that the traditional algorithms can be extend to discover the time-interval sequential patterns without pre-defining any time partitions. The result of our example verifies that the concept of our proposed algorithm is more reasonable and efficient.

**Authors**

Dr. Hao-En Chueh received his B.S. in Computer Science and Information Engineering from Tamkang University, Taiwan in 1999 and M.S. in Computer Science and Information Engineering from Tamkang University, Taiwan in 2001. He received his Ph.D. in Computer Science and Information Engineering from Tamkang University, Taiwan in 2007. His research areas include data dining, fuzzy theory, probability theory, statistics, database system and its applications. He is an assistant professor of the Department of Information Management at Yuanpei University, Hsinchu, Taiwan.

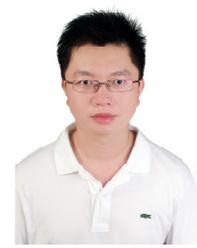